\pdfoutput=1

\documentclass[letterpaper,11pt,preprint]{aastex}
\usepackage{amssymb,charter,natbib}
\newcommand{\e}[1]{\times 10^{#1}}                        

\newcommand{\fig}[1]{Figure \ref{#1}}

\newcommand{\figwide}[3]{
   \begin{figure*}[!ht]
   \centering
   \includegraphics{#1}
   \caption{#3}
   \label{#2}
   \end{figure*}
}

\begin{document}

\title{Rayleigh-Taylor-Unstable Accretion and Variability of Magnetized Stars: Global Three-Dimensional Simulations}
\shorttitle{Unstable Accretion to Magnetized Stars and Variability}

\author{Akshay K. Kulkarni, Marina M. Romanova}
\affil{Dept. of Astronomy, Cornell University, Ithaca, NY 14853}
\email{akshay@astro.cornell.edu, romanova@astro.cornell.edu}


\begin{abstract}
We present results of 3D simulations of MHD instabilities at the accretion disk-magnetosphere boundary. The instability is Rayleigh-Taylor, and develops for a fairly broad range of accretion rates and stellar rotation rates and magnetic fields. It produces tall, thin tongues of plasma that penetrate the magnetosphere in the equatorial plane. The shape and number of the tongues changes with time on the inner-disk dynamical timescale. In contrast with funnel flows, which deposit matter mainly in the polar region, the tongues deposit matter much closer to the stellar equator. The instability appears for relatively small misalignment angles, $\Theta\lesssim30^\circ$, between the star's rotation and magnetic axes, and is associated with higher accretion rates. The hot spots and light curves during accretion through instability are generally much more chaotic than during stable accretion. The unstable state of accretion has possible implications for quasi-periodic oscillations and intermittent pulsations from accreting systems.
\end{abstract}


\section{Introduction}
The geometry of the accretion flow around magnetized stars is an important factor in determining the observed spectral and variability properties of the accreting system. An accretion disk around a magnetized central object is truncated at the distance from the central star where the magnetic energy density becomes comparable to the matter energy density. Beyond that point, the gas can accrete to the star (1) through funnel streams, or magnetospheric accretion \citep[e.g.,][]{GhoshLamb79}, or (2) through plasma instabilities at the disk-magnetosphere interface \citep[e.g.,][]{AronsLea76, ElsnerLamb77}. The geometry of the matter flow in these two regimes is expected to be very different. In general, the Rayleigh-Taylor (RT), or interchange, instability is expected to develop at the disk-magnetosphere interface because of the high-density disk matter being supported against gravity by the low-density magnetospheric plasma. The Kelvin-Helmholtz (KH) instability is also expected to develop because of the discontinuity in the angular velocity of matter at the boundary. The inner disk matter is expected to rotate at the local Keplerian velocity, while the magnetospheric plasma corotates with the star.

Two- and three-dimensional simulations have shown accretion through funnel streams \citep{RomanovaEtAl04, KulkarniRomanova05}. Here we report on accretion through the Rayleigh-Taylor, or interchange, instability in global 3D MHD simulations.


\section{Simulation Results}

Our numerical model and reference values are described in \citet{KulkarniRomanova08}. We chose the following parameters for our main case: dipole moment $\mu=2$, corresponding to an equatorial surface magnetic field of $B_*=2$, misalignment angle $\Theta=5^\circ$, viscosity parameter $\alpha=0.1$, stellar rotation period $P=3$, initial disk radius $=2$. The corotation radius, which is the radius at which the orbital rotation rate of the disk matter equals the star's rotation rate, is $r_{cor}=2$.

\figwide{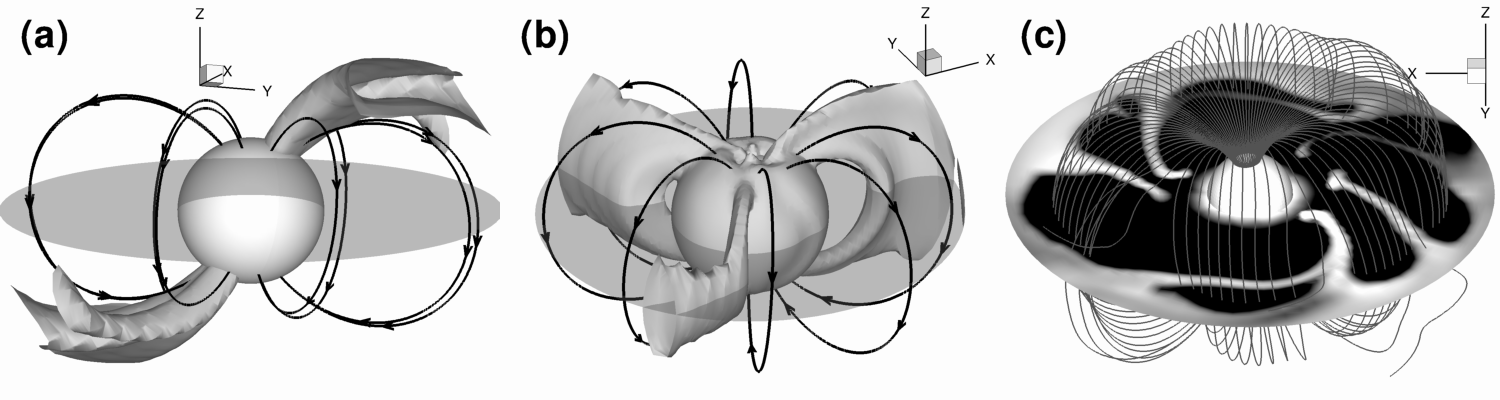}{3d}{Accretion through funnels (a) and instability (b). A constant density surface is shown. The lines are magnetospheric field lines. The translucent disc denotes the equatorial plane. The star's rotation axis is in the z-direction. (c) A tongue of gas, shown by density contours in the equatorial plane, pushing aside magnetic field lines on its way to the star.}

\fig{3d} compares the accretion flow in the stable and unstable cases. Unstable perturbations at the disk-magnetosphere boundary grow and disk matter penetrates the magnetosphere in the form of tongues of gas travelling through the equatorial plane by prying the field lines aside. Matter energy density dominates inside the tongues. When they come closer to the star, they encounter a stronger magnetic field, which stops their equatorial motion. At this point the tongues turn into miniature funnel-like flows following the field lines. They deposit matter much closer to the star's equator than true funnel flows do.

\figwide{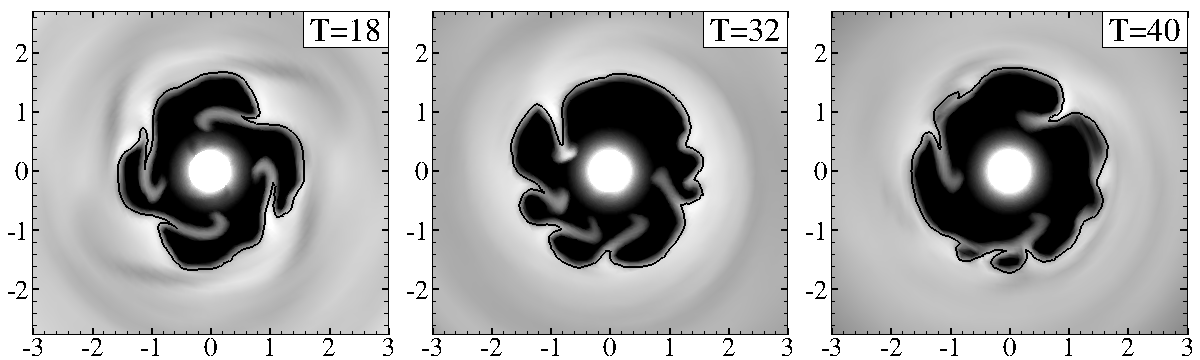}{xy}{Equatorial slices of the circumstellar region for our main case. The shades represent plasma density contours. The black line is the $\beta=1$ line.}

\fig{xy} shows equatorial slices of the circumstellar region at various times. The density enhancements which result in the formation of the tongues can be seen at the bases of the tongues. The number of tongues changes with time of its own accord, without any artifically introduced perturbation. The total number of tongues at any given time is of the order of a few.


\section{Effect of the Instability on Hotspots and Variability}
\fig{variab} compares the hot spots on the star's surface in the stable and unstable cases. We see that the spots are very different in the two cases. Each tongue creates its own hot spots when it reaches the star's surface. Therefore, the shape, intensity, number and position of the spots change on the inner-disk dynamical timescale.

\figwide{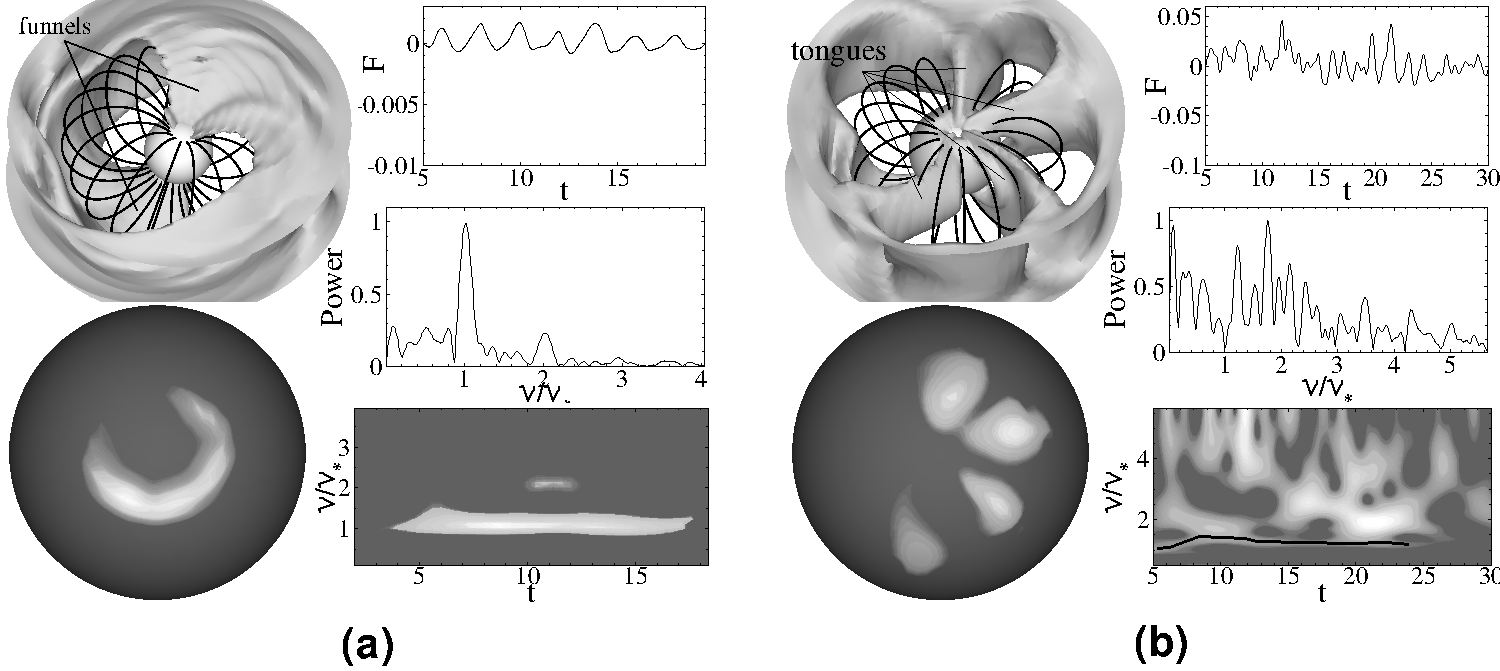}{variab}{(a) Accretion flow and hot spots (left column), and the lightcurve and its fourier and wavelet spectra (right column) in a stable accretion case, which has the following parameters: $\alpha=0.02$, $\Theta=15^\circ$, $P=2$, $\mu=2$. (b) The same for an unstable accretion case, which has the following parameters: $\alpha=0.2$, $\Theta=5^\circ$, $P=2.8$, $\mu=2$. The observer inclination angle is $i=45^\circ$ in both cases.}

We calculate lightcurves from the hotspots for various viewing directions, taking into account the general relativistic effects of light bending and gravitational redshift, and the Doppler effect \citep{PoutanenGierlinski03, KulkarniRomanova05}. The lightcurves show oscillations superimposed on a slowly varying background, which we subtract out. \fig{variab}a shows the lightcurve and its fourier and wavelet spectra for one of our stable accretion cases. We see a definite peak at the star's frequency. As the wavelet spectrum shows, there is no drift in the frequency. When the accretion is unstable, the tongues and the hotspots they produce rotate around the star's axis at a frequency different from the star's rotation frequency. Therefore, the star's frequency can disappear completely from the power spectrum, as seen for the visibly chaotic matter lightcurve in \fig{variab}b. A multitude of other peaks appears in the power spectrum instead. In this case, there are two prominent peaks with $\nu/\nu_* = 1.2$ and 1.8. There is also considerable drift in the frequency, as seen from the wavelet spectrum.


\section{Empirical conditions for the existence of the instability}
To investigate the parameter ranges over which the instability appears, we performed simulation runs for a variety of values of the accretion rate and and the star's rotation period $P$, for two magnetospheric sizes $r_m/R_*=$ 2-3 and 4-5, at misalignment angle $\Theta=5^\circ$. The accretion is unstable if

\begin{equation}
\dot M \ge \dot M_{*,crit} \approx x\,\, M_\odot \mbox{ yr}^{-1} \times
\left(\frac{B}{10^9\mbox{ G}} \right)^2 \left(\frac{R}{10\mbox{ km}} \right)^{5/2} \left(\frac{M}{1.4M\odot} \right)^{-1/2},
\nonumber
\end{equation}
where $x=1.8\e{-8}$ and $2.2\e{-9}$ for the small and large magnetospheres.


\section{Discussion}
Accretion through instability is expected to occur in most accreting systems for typical values of mass, radius, surface magnetic fields and accretion rates. One of the most interesting observational consequences of accretion through instabilities is the effect on the variability. Light curves associated with funnel accretion show clear pulsations at either the star's frequency or twice that, depending on the misalignment angle and the viewing geometry. In contrast, lightcurves associated with unstable accretion do not show very clear signs of periodicity. Since unstable accretion occurs at high accretion rates, this has a few implications: (1) If the accretion rate is close to the boundary between stable and unstable regimes, then slight changes in the accretion rate can cause the accretion to episodically switch between stable and unstable, causing corresponding appearance and disappearance of pulsations. This might be a possible explanation for the behaviour of intermittent pulsars \citep{AltamiranoEtAl08, CasellaEtAl08}. One of the attractive features of this idea comes from the fact that during stable accretion, the azimuthal location of funnels with respect to the star is fixed, even across periods of unstable accretion. The pulsations in intermittent pulsars would therefore be expected to be coherent in phase across periods of lack of pulsations. This ``phase memory'' has been observed in the intermittent pulsars mentioned above. (2) The stellar magnetic field is not buried even in the unstable cases that we have considered. So pulsations may be undetectable at even lower accretion rates than those required for field burial. (3) The fourier spectra of the unstable lightcurves do show some peaks. These appear because the number of tongues does not usually change very rapidly. If a certain number of tongues dominates for a significantly long time, it may lead to quasi-periodic oscillations in the lightcurves \citep{LiNarayan04}, which may be important for understanding Type II (accretion-driven) bursts in LMXBs.


\bibliography{ms}

\begin{thebibliography}{10}
\providecommand{\natexlab}[1]{#1}
\providecommand{\url}[1]{\texttt{#1}}
\expandafter\ifx\csname urlstyle\endcsname\relax
  \providecommand{\doi}[1]{doi: #1}\else
  \providecommand{\doi}{doi: \begingroup \urlstyle{rm}\Url}\fi

\bibitem[{Altamirano} et~al.(2008){Altamirano}, {Casella}, {Patruno},
  {Wijnands}, and {van der Klis}]{AltamiranoEtAl08}
D.~{Altamirano}, P.~{Casella}, A.~{Patruno}, R.~{Wijnands}, and M.~{van der
  Klis}.
\newblock {Intermittent Millisecond X-Ray Pulsations from the Neutron Star
  X-Ray Transient SAX J1748.9-2021 in the Globular Cluster NGC 6440}.
\newblock \emph{\apjl}, 674:\penalty0 L45--L48, February 2008.
\newblock \doi{10.1086/528983}.

\bibitem[{Arons} and {Lea}(1976)]{AronsLea76}
J.~{Arons} and S.~M. {Lea}.
\newblock {Accretion onto magnetized neutron stars - Structure and interchange
  instability of a model magnetosphere}.
\newblock \emph{\apj}, 207:\penalty0 914--936, August 1976.

\bibitem[{Casella} et~al.(2008){Casella}, {Altamirano}, {Patruno}, {Wijnands},
  and {van der Klis}]{CasellaEtAl08}
P.~{Casella}, D.~{Altamirano}, A.~{Patruno}, R.~{Wijnands}, and M.~{van der
  Klis}.
\newblock {Discovery of Coherent Millisecond X-Ray Pulsations in Aquila X-1}.
\newblock \emph{\apjl}, 674:\penalty0 L41--L44, February 2008.
\newblock \doi{10.1086/528982}.

\bibitem[{Elsner} and {Lamb}(1977)]{ElsnerLamb77}
R.~F. {Elsner} and F.~K. {Lamb}.
\newblock {Accretion by magnetic neutron stars. I - Magnetospheric structure
  and stability}.
\newblock \emph{\apj}, 215:\penalty0 897--913, August 1977.

\bibitem[{Ghosh} and {Lamb}(1979)]{GhoshLamb79}
P.~{Ghosh} and F.~K. {Lamb}.
\newblock {Accretion by rotating magnetic neutron stars. II - Radial and
  vertical structure of the transition zone in disk accretion}.
\newblock \emph{\apj}, 232:\penalty0 259--276, August 1979.
\newblock \doi{10.1086/157285}.

\bibitem[{Kulkarni} and {Romanova}(2005)]{KulkarniRomanova05}
A.~K. {Kulkarni} and M.~M. {Romanova}.
\newblock {Variability Profiles of Millisecond X-Ray Pulsars: Results of
  Pseudo-Newtonian Three-dimensional Magnetohydrodynamic Simulations}.
\newblock \emph{\apj}, 633:\penalty0 349--357, November 2005.
\newblock \doi{10.1086/444489}.

\bibitem[{Kulkarni} and {Romanova}(2008)]{KulkarniRomanova08}
A.~K. {Kulkarni} and M.~M. {Romanova}.
\newblock {Accretion to magnetized stars through the Rayleigh-Taylor
  instability: global 3D simulations}.
\newblock \emph{\mnras}, 386:\penalty0 673--687, May 2008.
\newblock \doi{10.1111/j.1365-2966.2008.13094.x}.

\bibitem[{Li} and {Narayan}(2004)]{LiNarayan04}
L.-X. {Li} and R.~{Narayan}.
\newblock {Quasi-periodic Oscillations from Rayleigh-Taylor and
  Kelvin-Helmholtz Instability at a Disk-Magnetosphere Interface}.
\newblock \emph{\apj}, 601:\penalty0 414--427, January 2004.
\newblock \doi{10.1086/380446}.

\bibitem[{Poutanen} and {Gierli{\'n}ski}(2003)]{PoutanenGierlinski03}
J.~{Poutanen} and M.~{Gierli{\'n}ski}.
\newblock {On the nature of the X-ray emission from the accreting millisecond
  pulsar SAX J1808.4-3658}.
\newblock \emph{\mnras}, 343:\penalty0 1301--1311, August 2003.
\newblock \doi{10.1046/j.1365-8711.2003.06773.x}.

\bibitem[{Romanova} et~al.(2004){Romanova}, {Ustyugova}, {Koldoba}, and
  {Lovelace}]{RomanovaEtAl04}
M.~M. {Romanova}, G.~V. {Ustyugova}, A.~V. {Koldoba}, and R.~V.~E. {Lovelace}.
\newblock {Three-dimensional Simulations of Disk Accretion to an Inclined
  Dipole. II. Hot Spots and Variability}.
\newblock \emph{\apj}, 610:\penalty0 920--932, August 2004.
\newblock \doi{10.1086/421867}.

\end{thebibliography}

\end{document}